\documentclass[aps,prl,twocolumn,superscriptaddress,showpacs,floatfix,preprintnumbers]{revtex4-1}

\usepackage[dvipsnames]{xcolor}     
\definecolor{lcolor}{rgb}{0.5,0,0}
\definecolor{citcolor}{rgb}{0,0,1}
\usepackage[breaklinks,colorlinks,urlcolor=blue,citecolor=citcolor,linkcolor=lcolor,linktoc=all]{hyperref}
\usepackage{color}
\usepackage{graphicx}	
\usepackage[utf8]{inputenc}
\usepackage{amsmath} \usepackage{amssymb}
\usepackage{bm,bbm,bbold}

\newcommand{\be}{\begin{equation}}
\newcommand{\ee}{\end{equation}}
\newcommand{\pa}{\partial}
\newcommand{\calR}{{\cal R}}

\newcommand{\deltaV}{\delta V}

\begin{document}
\title{A black hole effective theory for strongly interacting matter}

\preprint{APCTP Pre2024 - 012}

\author{Matti Järvinen}
\affiliation{Asia Pacific Center for Theoretical Physics, Pohang, 37673, Korea}
\affiliation{Department of Physics, Pohang University of Science and Technology, Pohang, 37673, Korea}
\author{Dorin Weissman}
\affiliation{Asia Pacific Center for Theoretical Physics, Pohang, 37673, Korea}

\begin{abstract}
We establish a new tool for studying strongly coupled matter: 
an effective theory of 
black holes in gravity, which maps to a hydrodynamic description of field theories via the gauge-gravity duality.
Our approach is inspired by previously known effective theories found in the limit 
of high number of dimensions.  We argue that the new approach can 
accurately describe phase transitions in a wide class of theories, such as the Yang-Mills 
and other nearly critical field theories. As an application, we analyze the 
interface between confining and deconfining phases in holographic Yang-Mills theory.
\end{abstract}

\maketitle
\section{Introduction}
Analyzing phase transitions and related dynamical phenomena in strongly coupled field theories is  a challenging problem with interesting applications. A prime example of a phase transition in a strongly interacting system is provided by  
the Yang-Mills theory, which has a first order confinement-deconfinement phase transition. In addition, full QCD may have a  first order transition at finite baryon number density, which can potentially be probed directly in heavy-ion collisions in the future.

First order phase transitions may also take place in the early universe.  
They produce gravitational waves, which are potentially observable. The standard model does not give rise to such phase transitions, but they could arise from strongly coupled extensions of the electroweak sector, 
strongly interacting 
theories at higher energies, or hidden sectors. Gravitational waves produced by the phase transitions near the electroweak scale may be observed by the LISA experiment~\cite{LISA:2022yao}, while transitions at much higher energies could explain the gravitational wave background recently observed by the NANOGrav experiment using pulsar timing~\cite{NANOGrav:2023gor}.  

Detailed studies of phase transitions and gravitational wave production require real-time dynamics, which are challenging to analyze directly in  strongly coupled field theories.
It is therefore useful to search for alternative approaches. 
A modern approach is to use the gauge-gravity duality, which maps the time-evolution of the strongly coupled system to that of higher dimensional classical gravity. This however does not remove all issues: For example, computing the time-evolution in a generic 3+1 dimensional setup in field theory, with  dependence on all spatial coordinates, requires solving full five-dimensional classical evolution on the gravity side, which is numerically extremely costly (see e.g.~\cite{Bellantuono:2019wbn,Bantilan:2020pay,Bea:2022mfb}).

The numerical analysis can be greatly simplified by considering gravity solutions in a high number of dimensions $D$, dual to conformal theories in $D-1$ dimensions \cite{Emparan:2013moa,Emparan:2020inr}. Finite temperature configurations are dual to black holes on the gravity side, and the dynamics of the black holes boils down to an effective theory having the same dimensionality as the field theory \cite{Emparan:2015gva,Emparan:2015hwa} (see also~\cite{Bhattacharyya:2015dva,Dandekar:2016jrp} and~\cite{Janik:2021jbq,Ares:2021nap,Ares:2021ntv,Janik:2022wsx,Fichet:2023dju} for complementary approaches). This effective theory is a nonlinear hydrodynamic theory which 
describes the membrane-like motion of the black hole horizon. The theory takes a non-relativistic form with linear time derivatives, which makes simulations straightforward and much easier than directly simulating Einstein gravity~\cite{Emparan:2015gva,Andrade:2018yqu,Emparan:2021ewh,Licht:2022rke,Luna:2022tgh,Emparan:2023dxm}. The approach is different from the fluid-gravity correspondence~\cite{Hubeny:2011hd} where one introduces an explicit derivative expansion: at large $D$, the hydrodynamic sector is decoupled from other modes~\cite{Emparan:2014aba}, such that expanding in $1/D$ automatically leads to a hydrodynamic theory.

Going to conformal theories at high number of dimensions seems to take one far away from the physically interesting 3+1 dimensional examples. This is however not  
the case: there is a link between the large $D$ approach and a class of lower-dimensional strongly coupled theories that include, in particular, the Yang-Mills theory (see~\cite{Betzios:2018kwn,Gursoy:2021vpu}). 
Specifically,   
the large $D$ gravity theory can be dimensionally reduced~\cite{Kanitscheider:2008kd,Gouteraux:2011qh} to five-dimensional dilaton gravity with a potential that is similar 
to that used in the improved holographic QCD (IHQCD) model, which is known to give a precise description of the Yang-Mills theory via the holographic duality \cite{Gursoy:2007cb,Gursoy:2007er}.  The same holds for spin-models near criticality~\cite{Gursoy:2010kw}. These latter models belong to a wider class of theories which have a higher order phase transition at finite temperature~\cite{Gursoy:2010jh}. 

The aim of this letter is to combine the ideas discussed above to derive explicitly a large-$D$-inspired hydrodynamic theory that can be readily applied to carry out lightweight dynamical simulations of strongly coupled matter in low-dimensional systems. In order to achieve this we first find a generalization of the previously known 
effective theories. It works not only for theories obtained by dimensionally reducing high-dimensional gravity, but also for their generic deformations. Importantly, this generalization allows one to study the physics of phase transitions and interfaces. Among others, it covers geometries found in the IHQCD model and duals of near-critical spin models. 
As an example of an application, we 
construct the interface between the confined and deconfined phases of the IHQCD model by using the new framework.

\section{The Large D inspired model}

Our approach is inspired by the large $D$ limit of general relativity \cite{Emparan:2013moa,Emparan:2015gva,Emparan:2015hwa,Andrade:2018zeb}. The typical starting point in this approach is 
Einstein gravity with a cosmological constant in $D=d+n$ dimensions. Considering a particular class of solutions with a non-trivial metric dependence only on the first $d$ dimensions~\footnote{For details, see the brief review provided in part I of the supplemental material.}, one can dimensionally reduce~\cite{Kanitscheider:2008kd,Gouteraux:2011qh} the high-dimensional gravity action to the Einstein-dilaton action,
\begin{equation} S = \frac{1}{16\pi G_5}\int d^5x \sqrt{-g}\left(R - \frac43(\pa \phi)^2 + V(\phi)\right) \,,\label{eq:EDaction} \end{equation}
where we have specialized to $d=5$, in which we will write our holographic model. 

In the limit $D\to\infty$ (equivalent to $n\to\infty$), the dilaton potential following from the dimensional reduction is $V(\phi)=V_0 e^{4\phi/3}$. Interestingly, this potential is exactly the critical potential separating deconfining and confining setups~\cite{Gursoy:2007cb}.
The dimensional reduction when $n$ is large but finite gives a specific correction to this potential. At this stage, our large $D$ ``inspired'' model takes a departure from the ordinary expansion. Instead of fixing the potential at large $n$ to that resulting from the dimensionally reduced pure gravity model, we consider a generic perturbation in $1/n$ around the $n\to\infty$ limit by taking
\begin{equation} 
 V(\phi) = V_0 e^{\frac43\phi}\left(1+\frac1n \delta V(\phi) + {\mathcal O}\left(\frac1{n^2}\right)\right) \,.\label{eq:potential}
\end{equation}
The perturbation $\delta V(\phi)$ is left as an arbitrary function of $\phi$. 
The constant $V_0$ is parametrized as
\begin{equation} V_0 = \frac{n^2}{\ell^2}\,. 
\label{eq:V0} \end{equation}
The explicit scaling with $n^2$ is necessary for the consistent expansion at large $n$. We see that the potential indeed agrees with the choices used for spin models in~\cite{Gursoy:2010jh,Gursoy:2010kw} at large values of $\phi$, corresponding to low temperatures. Moreover, the potential is close to that of the IHQCD model~\cite{Gursoy:2007cb,Gursoy:2007er} at large $\phi$.

\section{Black hole effective theory}
We can find non-trivial solutions to the Einstein-dilaton model \eqref{eq:EDaction} with potential \eqref{eq:potential} by constructing an effective theory in the $1/n$-expansion in the vein of~\cite{Emparan:2013moa,Emparan:2015gva,Emparan:2015hwa,Andrade:2018zeb}. 
To this end 
we write an ansatz for the metric in Eddington-Finkelstein coordinates, 
\begin{equation*} ds^2 = e^{2A}\left(-f dt^2 - 2\,dt\,dr - 2 C_i\, dt\, dx^i + g_{ij} dx^i dx^i\right) \,.
\end{equation*}
Including the first non-trivial corrections in $1/n$, the functions describing a generic evolving black hole are
\begin{align} \label{eq:Aansatz}
A &= -\frac{n r}{3\ell} + \frac 1n A^{(1)}({\cal R},t,x^k)\,, \\
f &= 1 - m(t,x^k) \calR + \frac{1}{n} f^{(1)}({\cal R},t,x^k)\,,
\\
g_{ij} &= \frac{1}{n}\delta_{ij} +\frac{1}{n^2}\frac{p_i(t,x^k) p_j(t,x^k)}{m(t,x^k)}{\cal R} \,,\label{eq:gijansatz}
\\
C_i &= \frac{1}{n} p_i(t,x^k) \calR 
+ \frac{1}{n^2}C_i^{(1)}(\calR,t,x^k) \,, \label{eq:gci}
\end{align}
where
\begin{equation}\label{eq:R} \calR \equiv e^{\frac{n r}{\ell}} \,. \end{equation}
The 
ansatz also includes the dilaton 
\begin{equation} \phi = \frac{n r}{2\ell} + \frac{1}{n}\phi^{(1)}(\calR,t,x^k)\,. \end{equation}
The space and time dependent functions $m$ and $p^i$ characterize the size and velocity of the black hole~\cite{Emparan:2015gva}, and the corrections $A^{(1)}$, $f^{(1)}$,  $C_i^{(1)}$, and $\phi^{(1)}$ are at this point arbitrary.
To leading order in $1/n$, the black hole horizon is located at ${\calR}_h = e^{\frac{n r_h}{\ell}} = 1/m$. The region outside it is $-\infty < r < r_h$, where $0 < {\cal R} < {\calR}_h$.

Some comments are in order.
Note that the zeroth order solution with constant $m$ is the same as the $n\to\infty$ solution (when $V=V_0 e^{4\phi/3}$ exactly) 
which matches with a well-studied string theory solution, the linear dilaton background~\cite{Aharony:1998ub,Bertoldi:2009yi,Gursoy:2010jh}. 
Moreover, 
the dependence of the metric on $r$ in~\eqref{eq:R} 
comes with an explicit factor of $n$. As in the large $D$ literature, we consider the large $n$ limit with $\calR$ remaining finite. Therefore the expansion is also a near-horizon limit, as $r - r_h = \frac{\ell}{n}\log(m\calR)$ is small, giving rise to a membrane-like picture.
The coordinates $x_i$ have also been rescaled to reflect the fact that the length scale of fluctuations in these directions is (before rescaling) of order $1/\sqrt{n}$. In the rescaled coordinates 
the expansions for $g_{ij}$ and $C_i$ start at order $1/n$.  This reflects the usual scalings found at large $D$~\cite{Emparan:2013moa,Emparan:2015gva}.

At this point we are ready to start solving the Einstein equations and the equation of motion for the dilaton as a series in $1/n$. We find explicit solutions for the corrections $f^{(1)}$, $A^{(1)}$, $C^{(1)}$ and $\phi^{(1)}$ in terms of $m$, $p_i$, and the potential $\delta V$, and then impose regularity at the black hole horizon~\footnote{Some technical subtleties regarding these solutions are discussed in part II of the supplemental material.}.

This leads to the following effective equations for $m$ and $p_i$:
\begin{align}
 \pa_t m &= \ell \pa_i \pa^i m -\pa_i p^i \,,
 \label{eq:effeq1} \\
 \pa_t p_i &= \ell \pa_j \pa^j p_i - \pa_j \left(\frac{p^j p_i}{m}\right) + \frac12 \delta V^\prime\left(-\frac12\log m\right)\pa_i m \,,
  \label{eq:effeq2}
\end{align}
which constitute the main analytic result of this letter.

The effective equations reduce exactly to those found in \cite{Emparan:2015gva} when taking $\delta V(\phi) = 2\phi$ (for flat space solutions reduced over spherical extra dimensions), or $\delta V(\phi) = -2\phi$ (for AdS solutions reduced over flat extra dimensions) \footnote{See part I of the supplemental material for details on the reduction and explicit expressions.}. 

\section{Thermodynamics and hydrodynamics}
Before solving the effective equations, we 
derive 
thermodynamic relations.  In this computation, we can simply set $p_i$ to zero and $m$ to a constant value, which needs to hold to a good approximation for the system to be near equilibrium.
The temperature is the Hawking temperature of the black hole, which can be determined in a standard way by computing its surface gravity, proportional to $df/dr$ at the horizon. 
Including the $\mathcal{O}\left(1/n\right)$ correction one finds \footnote{See part~II in the supplemental material for the derivation.} 
\begin{equation} T  
= \frac{n}{4\pi \ell}\left[1 + \frac1n \delta V\left(-\frac12 \log m\right)\right]\,. \label{eq:temperature} \end{equation}
The leading term is a constant, i.e., 
the black holes have the same temperature at any size in the $n\to\infty$ limit.
Interestingly, after including the correction, the temperature is proportional to the potential~\eqref{eq:potential},
\begin{equation} \label{eq:TfromV}
     T = \frac{\ell}{4\pi n} V(\phi)e^{-\frac43 \phi}\Big|_{\phi=-\frac12 \log m}
\end{equation}
with $\phi$ taking its zeroth order value at the horizon.
The entropy density is given by the area of the black hole,
\begin{equation} S
= \frac{1}{4G_5} m \,,\label{eq:entropy} 
\end{equation}
where we took into account the scaling of the spatial coordinates implied by eq. \eqref{eq:gijansatz}.

Let us comment on the perturbations of the black hole. Setting $m \mapsto m +e^{-i\omega t+iqx_1}\delta m $ and $p_i \mapsto  e^{-i\omega t+iqx_1}\delta p_i$, where the amplitudes of the perturbations $\delta m$, $\delta p_i$ are small, we solve the dispersion relations for the frequency $\omega$ and momentum $q$. Linearizing equations \eqref{eq:effeq1} and \eqref{eq:effeq2}, we find for perturbations $\delta p_i$ perpendicular to the momentum
\begin{equation*}
 \omega = - i \ell q^2 \ ,
\end{equation*}
whereas the perturbations  $\delta m$ and $\delta p_1$ give
\begin{equation*}
 \omega = \pm \sqrt{-\frac{\deltaV^\prime(-\frac{1}{2}\log m)}{2}}\, q - i \ell q^2 \ .
\end{equation*}
These modes are identified as the hydrodynamic modes, i.e., the dissipative shear modes and the sound modes, respectively. The speed of sound is related to the derivative of the potential. When the derivative becomes positive, there is a spinodal instability. Both these observations agree  with the thermodynamic relations \eqref{eq:temperature} and \eqref{eq:entropy}.

\section{Domain walls}
We now study 
domain wall solutions interpolating between two  
phases corresponding to black holes with different values of $m$. We focus on static solutions depending on only one spatial coordinate, $x\equiv x^1$.  
The first effective equation \eqref{eq:effeq1} tells us that $p^\prime = \ell m^{\prime\prime}$. 
We choose the solution with zero integration constant, $p = \ell m^\prime$, such that $p\to0$ in regions where $m$ goes to a constant. The remaining equation \eqref{eq:effeq2} is reduced to
\begin{equation} m^{(3)} - \frac{2 m^\prime m^{\prime\prime}}{m} + \frac{(m^\prime)^3}{m^2} = -\frac{1}{2\ell^2} m^\prime \delta V^\prime\left(-\frac12\log m\right)\,. \label{eq:effeqmx} \end{equation}
From this equation, we can derive two conditions for the existence of a domain wall solution, interpolating between $m(x\to-\infty) = m_0$ and $m(x\to\infty)=m_1$. The first is obtained by taking the integral $\int_{-\infty}^x dx^\prime$ of both sides of eq. \eqref{eq:effeqmx}:
\begin{equation} \label{eq:dweq1}
m^{\prime\prime} - \frac{(m^\prime)^2}{m} = -\frac{1}{2\ell^2}\int_{m_0}^{m(x)}\!\! dm\, \delta V^\prime\left(-\frac12 \log m\right)\,, \end{equation}
while the second is obtained by 
taking $\int_{-\infty}^x dx^\prime \frac{1}{m(x^\prime)}$:  
\begin{align} \label{eq:dweq2} &\frac{m^{\prime\prime}}{m}-\frac{(m^\prime)^2} {2m^2} = \\ \nonumber &\quad\frac{1}{\ell^2}\left[\delta V\left(-\frac12\log m(x)\right)-\delta V\left(-\frac12\log m_0\right)\right]\,. \end{align}
From these 
two equations it follows that, for a domain wall solution satisfying $m^\prime(\pm\infty) = 0 = m^{\prime\prime}(\pm\infty) $, 
\begin{equation} \int_{m_0}^{m_1} \!\! dm\, \delta V^\prime\left(-\frac12\log m\right) = 0 \label{eq:dwcon1} \end{equation}
and
\begin{equation} \delta V\left(-\frac12\log m_0\right) - \delta V\left(-\frac12\log m_1\right) = 0 \,.\label{eq:dwcon2}
\end{equation}

Using the relations~\eqref{eq:temperature}--\eqref{eq:entropy}, we can immediately translate eqs.~\eqref{eq:dwcon1} and~\eqref{eq:dwcon2}, respectively, into the conditions
\begin{equation} \Delta F = -\int\! S\, dT = 0  
\ , \qquad \Delta T = 0\,,\end{equation}
meaning that a stable domain wall solution can exist between configurations with the same free energy and temperature, consistently with thermodynamics.

Note that in the derivation of eq. \eqref{eq:dwcon2}, we assumed that $m_0$ and $m_1$ are both non-zero, which corresponds to a wall between two deconfined phases. A domain wall solution with one of the endpoints at zero (i.e., a wall between a confined and a deconfined phase~\cite{Aharony:2005bm,Figueras:2014lka}) 
does not have to satisfy the condition on the temperature, 
but only the 
condition on the free energy. This reflects the fact that a regular solution with $m=0$ exists at any temperature (see, e.g.,~\cite{Gursoy:2008bu,Gursoy:2008za}).
The domain wall solution constructed below is in this latter class.

\begin{figure*}
\includegraphics[width=\textwidth]{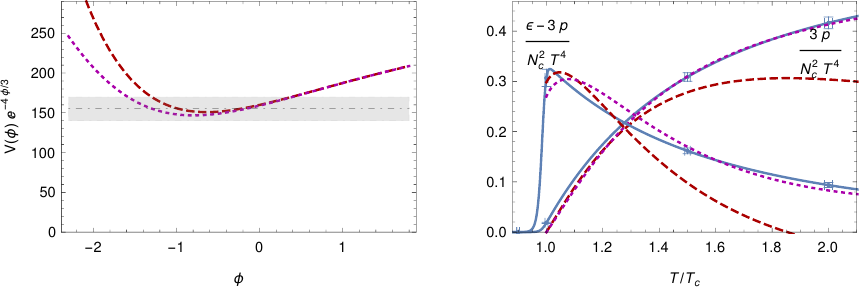}
    \caption{The IHQCD potentials used in this letter (left) and results from the large $D$ black hole thermodynamics compared to lattice data~\cite{Panero:2009tv} (right).  
    The red dashed (magenta dotted) curves show the potentials and thermodynamics for our first choice (second choice) of potentials. The gray band in the left panel shows the region where the deviation from the constant $\hat V_0$ is small, $|\delta V|<0.1$. In the right panel, the blue curves and error bars show the lattice results.}
    \label{fig:lattice_fit}
\end{figure*}

\section{Application to Yang-Mills theory}
As an application, we construct the domain wall between confined and deconfined phases in a holographic model for the Yang-Mills theory using our framework.
To this end, we choose the potential $V$ following the IHQCD model \cite{Gursoy:2007cb,Gursoy:2007er}. We consider two choices: 
\begin{enumerate}
    \item Using directly the IHQCD potential fitted using the full model (without invoking the expansion in $1/n$) to lattice data for thermodynamics and glueball spectrum in Yang-Mills theory.  
    That is, we will use the potential $V$ fitted in~\cite{Jokela:2018ers} (see also~\cite{Gursoy:2009jd}), and treat its deviation from the purely exponential form as the perturbation $\delta V$ in~\eqref{eq:potential}.
    \item A potential which is otherwise identical to the first choice, but modified close to the boundary (small values of $\phi$) such that it agrees better with the lattice data when the formulas~\eqref{eq:temperature} and~\eqref{eq:entropy} are used.
\end{enumerate}

Explicitly, the potentials are defined as 
\begin{align*} \label{eq:VIHQCD}
&V_\mathrm{IHQCD}(\phi) = 12\,\biggl[\,\frac{1}{1+\frac{c_\mathrm{UV}}{e^{\phi}}}+V_1 e^\phi+{V_2e^{2\phi}
\over 1+c_\mathrm{IR}e^\phi}+&\\
&+V_\mathrm{IR} \exp\left(-\frac{1}{c_\mathrm{IR}e^{\phi}}\right)e^{4\phi/3}c_\mathrm{IR}^{4/3}\sqrt{\log(1+c_\mathrm{IR}e^\phi)}\,\biggr]\,, &\nonumber
\end{align*}
where
\begin{equation*}
 V_1 = \frac{88}{27} \ , \quad V_2 = \frac{4619}{729} \ , \quad  c_\mathrm{IR} = 3 \ , \quad V_\mathrm{IR} = 2.05 \,,
\end{equation*}
for both choices. They differ only via the value of $c_\mathrm{UV}$:
\begin{align*} \nonumber
    c_\mathrm{UV} &= 0  &\quad &\textrm{(choice 1)}\,,&\\
    c_\mathrm{UV} &= 0.075  &\quad &\textrm{(choice 2)\,.}&
\end{align*}
In order to apply our approach we decompose the potential as in~\eqref{eq:potential}: 
\begin{equation*}
V_\mathrm{IHQCD}(\phi) = \hat V_0 e^{4\phi/3}\left[1+\delta V_\mathrm{IHQCD}(\phi)\right] \,,
\end{equation*}
where $\hat V_0 = 155$ was chosen such that $\delta V_\mathrm{IHQCD}$ is small in the intermediate region where $e^\phi = \mathcal{O}(1)$, which is most important for the phase transition. This is the region, where $V(\phi)e^{-4\phi/3}$, which maps to the temperature via~\eqref{eq:TfromV}, is near its minimum (see Fig.~\ref{fig:lattice_fit}, left panel). 
Note that we do not introduce a formal expansion parameter $n$ here, but since $\hat V_0$ is numerically large and $\delta V_\mathrm{IHQCD}$ is numerically small in the relevant region (the gray band in Fig.~\ref{fig:lattice_fit} indicates where $\delta V < 0.1$), our 
approach is expected to work well. 

The thermodynamics computed from~\eqref{eq:temperature} and~\eqref{eq:entropy} is compared to lattice data for 
Yang-Mills at large number of colors~\cite{Panero:2009tv} in Fig.~\ref{fig:lattice_fit} (right).
We analyze the pressure $P$ and the interaction measure $(\epsilon-3P)/T^4$, where $\epsilon$ is the energy density, and fit Newton's constant $G_5$ independently for each potential. As expected, choice~1 which was fitted to lattice data in the IHQCD model without approximations, produces a fair description of the data only near $T=T_c$ where $\delta V$ is small. Thanks to the adjustment close to the boundary, choice 2  gives a good fit in a larger range of temperatures. 

\begin{figure}
    \includegraphics[width=0.44\textwidth]{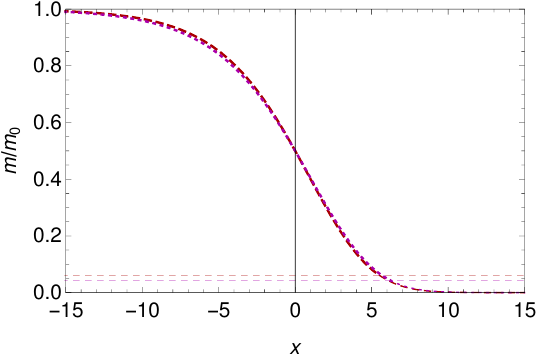}%
    \caption{The profiles of the domain walls shown as the red dashed (magenta dotted) curves for our first choice (second choice) of potentials. Note that the curves almost overlap. The horizontal dashed lines indicate the value of $m$ under which $|\delta V|$ becomes larger than 0.1.}
    \label{fig:domain_walls}
\end{figure}

The thermodynamics arise from the region of negative $\phi$ with $\deltaV^\prime(\phi)<0$, where the black hole solutions are stable. Therefore the difference between the potentials can be used to estimate the accuracy of the $1/n$ expansion in  this region:  
we show the domain wall solutions obtained by solving eqs.~\eqref{eq:dweq1} and~\eqref{eq:dweq2} for both potential choices in Fig.~\ref{fig:domain_walls}. Despite the difference between the potentials at negative $\phi$ shown in Fig.~\ref{fig:lattice_fit} (left), the solutions are nearly identical. 

At positive $\phi$ we do not have a direct handle to check the accuracy of the $1/n$ expansion, but we note that the potential exits the gray band of Fig.~\ref{fig:lattice_fit} (left) only in the tail of the domain wall (the part below the dashed lines in Fig.~\ref{fig:domain_walls}), which suggests that most of the domain wall is well approximated. That is, even though the range of $\phi$ where $\delta V$ is small and the potential remains inside the gray band in Fig.~\ref{fig:lattice_fit} (left) appears to be limited, almost all of the domain wall solution is described by this region. This happens because in the effective theory $\phi$ is replaced by $-\frac{1}{2}\log m$, a slowly varying function.

\section{Conclusions}

We derived a novel generalized black hole effective theory which can be applied to describe strongly interacting matter in a wide class of theories via the gauge-gravity duality. We pointed out that the generalized theory can describe dynamics of phase transitions and interfaces.
In particular,  we argued that 
it can be used to model Yang-Mills theory at finite temperature, near the phase transition. 

Our approach has some limitations. Firstly, it does not give exact solutions to the full five-dimensional gravity theory. It only captures the hydrodynamic modes, and not the time-dependence related to higher quasinormal modes of the black hole. However, in our framework, similarly to the large $D$ limit,
the hydrodynamic modes are decoupled from the rest~\cite{Emparan:2013moa,Emparan:2014aba} so that only including the hydrodynamic sector is a good approximation. 
Secondly, due to the non-relativistic form of our approach it is unlikely to be a good approximation for the Yang-Mills theory (or quark-gluon plasma) at temperatures well above the critical temperature. As the temperature grows, so does the speed of sound and the plasma is better approximated as a nearly conformal fluid.

The effective theory has several 
further applications, in particular to time-dependent scenarios. In the context of first order phase transitions in the early universe, it would be interesting to simulate moving domain walls or expanding bubbles and their collisions, and how they produce gravitational waves. Note that the approach includes crucial ingredients for gravitational wave production, e.g., sound waves. Other related problems would be to study the interfaces with higher order phase transitions, and the evolution for an interface between thermal baths at different temperatures~\cite{Bhaseen:2013ypa,Pourhasan:2015bsa,Ecker:2021ukv,Bachas:2021tnp}.

Another direction would be to improve and extend the approach further. First one could verify that the domain wall found in this letter is indeed a good approximation to the exact wall by solving it directly in IHQCD, which is a challenging but doable computation. 
More generally, one should check how the energy-momentum tensor is extracted. Doing this properly typically requires matching of the effective theory with behavior near the boundary,
as has been analyzed in the large D literature (see e.g.~\cite{Suzuki:2015axa,Emparan:2018bmi}).
One can also extend the approach systematically to higher order corrections in $1/n$, though there is no guarantee that these corrections can be solved analytically. Moreover, it might be possible to find similar effective theories in more general setups such as charged black holes~\cite{Bhattacharyya:2015fdk,Emparan:2016sjk}, which means going to finite density on the field theory side.


\acknowledgments

\emph{Acknowledgments.}---%
We thank U.~G\"ursoy, T.~Ishii, R.~Janik, N.~Jokela, G.~Policastro, E.~Pr\'eau, J.~Sonnenschein, J.~Subils, and B.~Withers for discussions. MJ has been supported by an appointment to the JRG Program at the APCTP through the Science and Technology Promotion Fund and Lottery Fund of the Korean Government and by the Korean Local Governments -- Gyeong\-sang\-buk-do Province and Pohang City -- and by the National Research Foundation of Korea (NRF) funded by the Korean government (MSIT) (grant number 2021R1A2C1010834). DW was supported by an
appointment to the YST Program at the APCTP through the Science and
Technology Promotion Fund and Lottery Fund of the Korean Government.
This was also supported by the Korean Local Governments --
Gyeongsangbuk-do Province and Pohang City.

\bibliography{References.bib}

\clearpage
\onecolumngrid
\appendix
\centerline{\bf \Large Supplemental material}
\setcounter{equation}{0}
\renewcommand{\theequation}{I.\arabic{equation}}

\section{I. Dimensional reduction}
We discuss here some details of dimensional reduction~\cite{Gouteraux:2011qh} from high-dimensional Einstein gravity to a dilaton gravity in lower number of dimensions. Starting from the Einstein-Hilbert action with a cosmological constant in $D = d + n$ dimensions,
\be S_{D} = \int d^d x\, d^n y\sqrt{-G}\left(R + \Lambda\right)\,, \ee
one can consider warped product spaces having a metric of the form
\be ds^2 = G_{\mu\nu}(x) dx^\mu dx^\nu + F^2(x) h_{ab}(y) dy^a dy^b \ee
where $\mu$, $\nu = 0,1,\ldots,d-1$ and $a$, $b = 1,2,\ldots,n$. Note that we do not assume that $n$ is large at this stage.

It is a straightforward exercise in general relativity to compute the Ricci curvature tensor of the $D$-dimensional space:
\begin{align} \label{eq:rmn}
R_{\mu\nu} &= R^{(G)}_{\mu\nu} - \frac{n}{F}\nabla_\mu \nabla_\nu F \,, \\ \label{eq:rab}
R_{ab}     &= R^{(h)}_{ab} - g^{\mu\nu}\left(\frac{1}{F}\nabla_\mu\nabla_\nu F + \frac{n-1}{F^2}\pa_\mu F \pa_\nu F\right)F^2 h_{ab} \,. \\
R_{a\mu}   &= 0 \,,
\end{align}
in terms of $R^{(G)}_{\mu\nu}$ and $R^{(h)}_{ab}$, which are the curvature tensors computed for the lower dimensional spaces described by the metrics $G_{\mu\nu}$ and $h_{ab}$ when considered separately and by themselves. Likewise, the covariant derivative $\nabla_\mu$ is that associated with the metric $G_{\mu\nu}$ alone.

The reduction becomes much simpler in the case where $h_{ab}$ is a space of constant curvature, and
\be R^{(h)}_{ab} = \frac{s (n-1)}{\ell^2} h_{ab}\,,\qquad R^{(h)} = \frac{s n(n-1)}{\ell^2}\,, \ee
where $s=0$ for flat space, $s=1$ for the sphere, and $s=-1$ for hyperbolic space. 
Then, equipped with the above expressions, one can write the Einstein equations in the full space in terms of only the metric $G_{\mu\nu}$ and the field $F(x)$.

Let us now introduce the dilaton field $\varphi(x)$ by writing the metric as
\be ds^2 = e^{2\delta_1 \varphi(x)} g_{\mu\nu}(x)dx^\mu dx^\nu + e^{2\delta_2\varphi(x)}h_{ab}dy^a dy^b\,. \ee
We have changed nothing except notation, writing
\be G_{\mu\nu} = e^{2\delta_1\varphi(x)} g_{\mu\nu}\,,\qquad F(x) = e^{\delta_2 \varphi(x)}\,. \ee
With the specific choice of
\be \delta_1 = \frac{\sqrt{n}}{\sqrt{(d-2)(n+d-2)}}\,,\qquad \delta_2 = -\frac{1}{\delta_1(n+d-2)} = -\frac{\sqrt{d-2}}{\sqrt{n(n+d-2)}}\,, \ee
the Einstein equations in $D=d+n$ dimensions are reduced to:
\be R_{\mu\nu} - \frac12 g_{\mu\nu} R - \frac12 g_{\mu\nu} V(\varphi)+\frac12 g_{\mu\nu}(\pa\varphi)^2 - \pa_\mu\varphi \pa_\nu \varphi = 0\,, \ee
\be \Box \varphi + \frac12V^\prime(\varphi) = 0 \,,\label{eq:boxphi} \ee
where the covariant derivatives and curvature tensor are computed now solely using the $d$-dimensional metric $g_{\mu\nu}$. These equations can be arrived at by taking eqs. \eqref{eq:rmn} and \eqref{eq:rab} and subjecting them to the conformal transformation of the metric $G_{\mu\nu}\to e^{2\delta_1\varphi}g_{\mu\nu}$. The coefficients $\delta_1$ and $\delta_2$ are chosen such that there are no additional terms in the dilaton equation of motion.

The dilaton potential obtained from the reduction, i.e the one defined by eq. \eqref{eq:boxphi}, is
\be V(\varphi) = -2\Lambda e^{2\beta\varphi} + \frac{s n(n-1)}{\ell^2} e^{2\beta^\prime\varphi}\,,
\ee
with
\be \beta = \delta_1\,,\qquad \beta^\prime =\frac{n+d-2}{n} \delta_1\,. \ee

Let us now use set $d=5$ and change the normalization of the dilaton by defining $\phi \equiv \sqrt{\frac43}\varphi$. The equations then are
\be R_{\mu\nu} - \frac12 g_{\mu\nu} = \frac12 g_{\mu\nu} V(\phi) - \frac43\left(\frac12g_{\mu\nu}(\pa\phi)^2-\pa_\mu\phi\pa_\nu\phi\right) \ee
and
\be \Box \phi + \frac38 V^\prime(\phi) = 0\,, \ee
with
\be V(\phi) = -2\Lambda e^{\frac43 x \phi} + \frac{s n(n-1)}{\ell^2}e^{\frac43 x^\prime\phi} \,.\label{eq:Vcompact} \ee
Here $x = \sqrt{3}\beta$ and  $x^\prime = \sqrt{3}\beta^\prime$.

These are exactly the equations of motion that follow from the reduced Einstein-dilaton action, which was the starting point of the model  in this letter:
\be S = \int d^5 x\sqrt{-g}\left(R - \frac43(\pa\phi)^2 + V(\phi)\right)\,. \ee

The potential dictated by the dimensional reduction has two exponential terms. The exponents depend on $n$ and $d$. If we take (for the first time in this calculation) that $n$ is large, the exponents are given by the expansion
\be x = \sqrt3\beta = 1 - \frac{3}{2n} + \ldots\,,\qquad x^\prime =\sqrt3\beta^\prime = 1 + \frac{3}{2n} +\ldots\,.
\ee
Expanding the potential itself at large $n$ will introduce linear (in $\phi$) correction terms at leading order, as:
\be \label{eq:Vcorrection} e^{\frac43x\phi} = e^{\frac43\phi}\left(1-\frac{2\phi}{n}+\ldots\right)\,, \qquad e^{\frac43x^\prime\phi} = e^{\frac43\phi}\left(1+\frac{2\phi}{n}+\ldots\right)\,. \ee

The two exponents in the potential are generated by two different sources. In the $n\to\infty$ limit the two exponents coincide and both give $e^{4/3\phi}$, but they differ when considering $1/n$ corrections. One term comes from the cosmological constant in the full $D$-dimensional action, and the second term is sourced by the curvature of the ``inner'' space spanned by the coordinates $y_a$. We can set either of these independently to zero, by setting $\Lambda=0$ (flat space) and taking the inner space to be curved, or by keeping $\Lambda\neq0$ and setting $h_{ab}=\eta_{ab}$.

This is exactly how we recover the results of the effective theory of~\cite{Emparan:2015gva} from our eqs.~\eqref{eq:effeq1}--\eqref{eq:effeq2}. In the former case we recover the flat space theory where $\Lambda=0$, the inner space is spherical ($s=1$), and we get the potential with $\deltaV = 2\phi$. The latter possibility yields the AdS effective theory, when $\Lambda<0$, the inner space is flat ($s=0$), resulting in $\deltaV = -2\phi$.

\setcounter{equation}{0}
\renewcommand{\theequation}{II.\arabic{equation}}

\section{II. Details on the effective theory and the large \texorpdfstring{$n$}{n} solution}
Solving the Einstein equations, plus the equation of motion for the dilaton, for the ansatz given in the main text starting from eq.~\eqref{eq:Aansatz}, at leading order in $1/n$ is in itself a straightforward task. There are some technical subtleties, however, which we would like to address here for the interested reader.

The first effective equation~\eqref{eq:effeq1} is obtained immediately from the $rt$ component of the Einstein equations. Equation~\eqref{eq:effeq2} comes from the $rx^i$ component, but to reach it one must first solve for the other corrections $A^{(1)}$, $f^{(1)}$, and $\phi^{(1)}$, as well as impose certain boundary and regularity conditions on their solutions. The corrections are given by lengthy expressions involving the functions $m$, $p_i$ and $\deltaV$.

The dependence of the solutions on the potential is through certain integral expressions. In particular there are the functions
\be I_V^{(1)}(\calR) = \int^\calR \frac{\delta V^\prime( + \frac12\log\rho)}{\rho^2}d\rho \label{eq:IV1} \ee
and
\be I_V^{(2)}(\calR,t,x^i) = I_V^{(2)}(\calR,m) = \int^\calR \frac{I_V^{(1)}(\rho)}{1-m \rho}d\rho \label{eq:IV2}\,. \ee
The latter expression has a logarithmic singularity at the horizon $m \calR = 1$, where it becomes $\approx I_V^{(1)}\left(\frac{1}{m}\right)\log(1 - m \calR)$. This singularity is canceled by choosing the coefficients $c(t,x^i)$ of terms which explicitly have the form $c(t,x^i)\log(1-m\calR)$ and appear in the solution to be exactly such that they will cancel the singularity of $I_V^{(2)}$. These $c(t,x^i)$ are integration constants left after integrating some equations with respect to ${\calR}$, and are set by this regularity condition. We have checked that neither the components of the metric nor the scalar curvature suffers from any singularities at the horizon with the appropriate choice of coefficients.

In addition to solving the Einstein equations, imposing regularity at horizon, we must fix some remaining integration constants by taking boundary conditions at ${\calR}\to 0$ ($r\to-\infty$), such that at $A$, $\phi$ and $f$ remain constant (i.e. have no $t$ or $x^i$ dependence) far away from the black hole.

Only after fixing these conditions do we have equation \eqref{eq:effeq2}, as well as the well defined temperature of eq. \eqref{eq:temperature}.

Note that one must also control the behavior of the integral functions for $\calR\to0$. We have written them as indefinite integrals, but the most precise way would be to define them as the definite integral from some finite cutoff $\epsilon$, taking $\int_\epsilon^\calR d\rho$, adding an appropriately chosen $\epsilon$-dependent regulator to eqs. \eqref{eq:IV1}--\eqref{eq:IV2} when necessary, and then taking the limit $\epsilon\to0$ to obtain the solution. The potential $\deltaV$ must be such that no singularities are introduced by these integral expressions, and we have a well defined solution at $\calR = 0$. For a given potential, one can evaluate these integrals explicitly or simply check their behavior as $\phi = \frac12\log\calR \to -\infty$ to verify their validity. A simple example is a polynomial correction $\deltaV = \sum_{k=1}^{k_{\mathrm{max}}} v_k \phi^k$ for which it is easy to get explicit expressions for the required integrals, and see that the solutions are well defined at $\calR\to0$.

We conclude by remarking on the derivation of the black hole temperature, eq.~\eqref{eq:temperature}. Including the $\mathcal{O}(1/n)$ correction, the horizon 
is generally located at
\begin{equation} {\cal R}_h(t,x^i) = \frac{1}{m(t,x^i)}\left[1+\frac1n f^{(1)}\left(\calR =\frac{1}{m(t,x^i)},t,x^{i}\right) \right]\,. \end{equation}
Using this we find that the correction to the black hole temperature, which is proportional to the derivative $df/dr$ at the horizon, at leading order is given by the expression
\be \label{eq:DeltaT} \Delta T = \frac{1}{\ell}\left(1 - \frac{1}{m}\frac{\pa}{\pa\calR}\right)f^{(1)}(\calR)\bigg\vert_{\calR=\frac{1}{m}} 
\ee
where we need to take $p_i=0$ for the stationary (constant $m$) solution. 

In order to find the temperature, one can insert in~\eqref{eq:DeltaT} the explicit and general space- and time-dependent solution for $f^{(1)}$ discussed above.
There however exists a much simpler derivation, 
which arises from the Einstein equations directly, and does not require solving the $1/n$ corrections explicitly. To be precise, solving $\phi^{(1)}$ from the $rt$ component, setting the integration constant in this solution to zero as required by boundary conditions at $\calR \to 0$, and inserting to the constraint equation (the $rr$ component) gives directly
\be \frac{1}{\ell}\left(1 - \frac{1}{m}\frac{\pa}{\pa\calR}\right)f^{(1)}(\calR)\bigg\vert_{\calR=\frac{1}{m}} = \deltaV\left( - \frac12\log m\right)\,. 
\ee
That is, the correction to the temperature turns out to be equal to the correction to the potential evaluated at the zeroth order location of the horizon $\calR = m^{-1}$. 
\end{document}